\documentclass[prl,twocolumn,showpacs,amssymb]{revtex4}
\usepackage{amsmath}
\usepackage{pdfpages, epsfig}

\begin{document}

\title{Dynamics of pattern-loaded fermions in bichromatic optical lattices}

\author{Matthew D. \surname{Reichl}}
	\affiliation{Laboratory of Atomic and Solid State Physics, Cornell University, Ithaca, New York 14853, USA}
	
\author{Erich J. \surname{Mueller}}
	\affiliation{Laboratory of Atomic and Solid State Physics, Cornell University, Ithaca, New York 14853, USA}

\date{\today}

\pacs{72.15.Rn, 37.10.Jk, 67.85.-d}

\begin{abstract}        % give a summary of your paper
%                         please supply keywords within your abstract
Motivated by experiments in Munich (M. Schreiber et. al. Science \textbf{349}, 842), we study the dynamics of interacting fermions initially prepared in charge density wave states in one-dimensional bichromatic optical lattices. The experiment sees a marked lack of thermalization, which has been taken as evidence for an interacting generalization of Anderson localization, dubbed ``many-body localization". We model the experiments using an interacting Aubry-Andre model and develop a computationally efficient low-density cluster expansion to calculate the even-odd density imbalance as a function of interaction strength and potential strength. Our calculations agree with the experimental results and shed light on the phenomena. We also explore a two-dimensional generalization. The cluster expansion method we develop should have broad applicability to similar problems in non-equilibrium quantum physics.

\end{abstract}
\maketitle

\begin{figure}  \vspace{-1.5em}
\hbox{\hspace{-3.9em}
\includegraphics[width=0.6\textwidth]{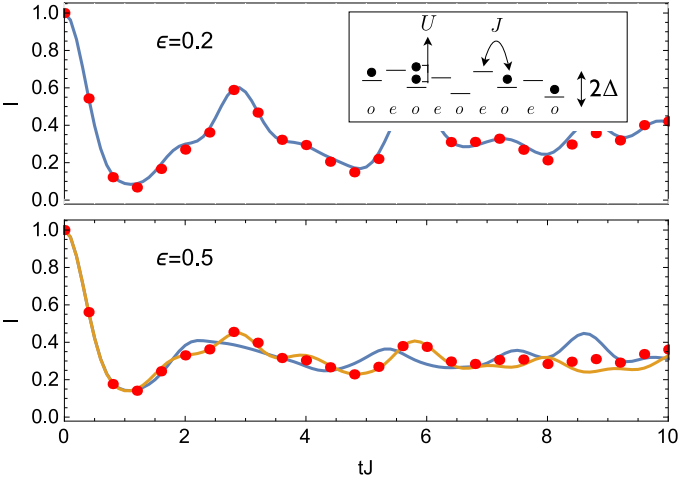}}
\caption{(Color online) Imbalance $I= \frac{N_{\rm{odd}}-N_{\rm{even}}}{N_{\rm{odd}}+N_{\rm{even}}}$ vs time $t$, measured in units of the nearest-neighbor hopping strength $J$ for fermions in an incommensurate superlattice of strength $\Delta$. $N_{\rm{odd}/\rm{even}}$ is the number of fermions on odd/even sites. The inset shows the geometry. At time $t=0$, $I=1$. The dark (blue) curves show the result of keeping the first two terms in the cluster expansion in Eq.~(\ref{eqifinal}) for 20 sites. The light (orange) curve shows the result of including three-particle terms in the cluster expansion. Red dots correspond to a time-dependent DMRG simulation. Here $\Delta = 3J$, $U=3J$, the superlattice period $\beta^{-1}=(0.721)^{-1}$ and the superlattice phase $\phi=0$.  The density is $\epsilon=0.2$ in the top graph and $\epsilon=0.5$ in the bottom graph.} 
\label{modelfig}
\end{figure}

%The inset shows a diagram of the physical setup: there is a pattern loaded lattice with nearest neighbor hopping $J$, incommensurate superlattice potential of strength $\Delta$, and on-site interaction $U$; at time $t=0$, fermions are placed randomly on only odd lattice sites and at later times the imbalance of the density on odd and even sites is measured. The main graphs show the  imbalance as a function of time for a lattice of 20 sites calculated using the cluster expansion in Eq.~(\ref{eqifinal}) (blue curves) and using time-dependent DMRG simulations (red points). %

%New techniques in controlling and imaging ultracold atoms has allowed unprecedented experimental access to the study of non equilibrium dynamics in closed quantum systems  \cite{polkovnikov2011}. Much of the experimental work in this field has been directed towards understanding how controlled disorder affects transport in these systems. 

\textit{Introduction} - An important challenge in many-body physics is to understand how interactions and disorder influence the transport properties of an electron gas. The non-interacting disordered problem was largely solved by Anderson \cite{anderson1958, abrahams1979}. By studying the expansion dynamics of wave packets of weakly interacting atoms, cold atom experiments have found evidence for Anderson localization in 1D \cite{billy2008} and 3D \cite{kondov2011, jendrzejewski2012} random speckled potentials and in 1D quasi random optical superlattices \cite{roati2008}. More recently, attention has turned to the interacting problem \cite{shepelyansky1996, barelli1996, eilmes1999, gornyi2005, basko2006, dufour2012,tezuka2012, iyer2013, serbyn2013, serbyn2014, huse2014, vosk2014, altman2015, li2015, modak2015, wang2015, nandkishore2014, eisert2015, devakul2015}. Schreiber et. al  \cite{schreiber2015}  devised an ingenious experiment to test these ideas. Here we model that experiment. 

%More recently, an experiment by Schreiber et. al  \cite{schreiber2015} studied transport dynamics and localization in 1D quasi-random superlattices by pattern loading interacting fermions into highly energetic initial states (in this case, charge density wave states) and probing their evolution at long times. Motivated by their work, we develop a cluster-expansion technique to theoretically study the dynamics of pattern-loaded fermions in optical lattices.%

The experiment in Ref.~\cite{schreiber2015} uses lasers to create a one-dimensional lattice with a weak periodic superlattice that is incommensurate with the main lattice (see the inset in Fig.~\ref{modelfig}). The resulting quasi-periodic potential shares features with a disordered one. For example, when the potential is sufficiently strong, all single particle states are localized. The experimentalists load interacting spin-$1/2$ fermions into some of the odd sites of the lattice, leaving the even sites empty. Some odd sites are doubly occupied. The atoms hop and interact for time $t$. The experimentalists measure the sublattice imbalance $I(t)$ 
\begin{equation}
I(t)= \frac{N_{\rm{odd}}-N_{\rm{even}}}{N_{\rm{odd}}+N_{\rm{even}}}
\end{equation}
where $N_{\rm{odd}/\rm{even}}$ is the number of fermions on odd/even sites at time $t$. In a localized phase, the atoms do not travel far from their initial position, and have a relatively high probability of being found at their starting point. Consequently in such a phase, one expects $I(t)$ to be non-zero at long times. Conversely, in a delocalized phase, one might expect $I(t)$ to decay to zero at long times. The experiment explores the long time behavior of $I$ as a function of superlattice strength and the interaction strength. The initial configuration of fermions on odd sites is random and the measurements are the result of ensemble averages over initial states. The experimentalists find two phases: one in which $I$ decays to zero, the other in which it is finite. The boundary appears to depend on the interactions in a non-monotonic manner.

In this paper we model the experiment, addressing the fundamental question of the interplay of incommensurate potentials and interactions. We develop a low-density cluster expansion which expresses the ensemble averaged imbalance as the sum of terms which involve only single-particle and two-particle dynamics. Using this computationally efficient approximation, we numerically calculate the long time imbalance as a function of interaction strength and superlattice strength. Our calculations reproduce the experimental results and provide insight into localization in the interacting system. We also extend our method to the case of a two dimensional lattice with an incommensurate superlattice in only one direction. The extra transverse degrees of freedom give kinetic pathways for equilibration; we calculate the consequences.

%\subsection{Model}%
\textit{Model and Methods} -  We model the atomic dynamics via the interacting Aubry-Andre model, given by the Hamiltonian  \cite{aubry1980, iyer2013}
\begin{equation} \label{hamileq}
\begin{split}
H= &-J \sum_{i, \sigma} \left(c^\dagger_{i, \sigma} c_{i+1, \sigma} + \mbox{h.c} \right)  \\
& + \Delta \sum_{i, \sigma}  \cos(2\pi \beta i +\phi) c^\dagger_{i, \sigma} c_{i, \sigma} + U \sum_i n_{i, \uparrow}n_{i, \downarrow}
\end{split}
\end{equation}
The first term describes nearest neighbor tunneling with strength $J$ while the second term describes a periodic superlattice potential of strength $\Delta$. For nearly all irrational values of $\beta$, this potential functions as quasi-random disorder which localizes all single particle states for sufficiently large superlattice strength ($\Delta /J  >2$) \cite{aubry1980}. In this regime, and for infinitely large systems, the single particle states are localized with a localization length $\lambda= (2\log \frac{\Delta}{2 J})^{-1}$, independent of $\beta$ \cite{aubry1980,sokoloff1985}. If $\beta= p/q$ is rational, the eigenstates are extended Bloch waves with period $q$. For large $\Delta$ and large $q$, the wavefunction in each unit cell is sharply peaked, and locally the eigenstates are similar to the irrational case.

\begin{figure}  \vspace{1.0em}
\hbox{\hspace{-1.8em}
\includegraphics[width=0.5\textwidth]{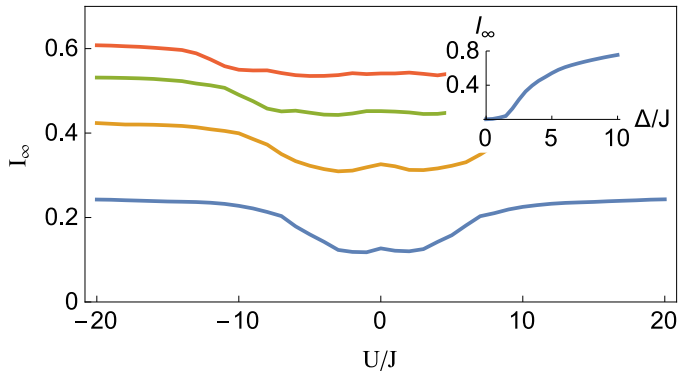}}
\caption{(Color online) Long time density imbalance $I_\infty$ as a function of interaction strength $U/J$ for a one-dimensional lattice with 20 sites at density $\epsilon=0.2$. The superlattice period is $\beta^{-1}=(0.721)^{-1}$ in units of the lattice spacing. The different curves correspond to different superlattice strengths: $\Delta/J=2,3,4,5$ (from bottom to top). The inset shows $I$ as a function of superlattice strength for $U/J=0$. } \label{fig2} 
\end{figure}

The localization transition is reflected in the observable $I(t)$, which for typical irrational $\beta$ and $U=0$ relaxes to $0$ for $\Delta/J<2$ but remains finite at long times for $\Delta /J >2$ (see the inset in Fig.~\ref{fig2}). We define $I_\infty = I(t \to \infty)$. Although $I_\infty \to 0$ as $\Delta/J \to 2$, the way it vanishes depends strongly on $\beta$ and is inconsistent with the naive estimate from structureless exponentially localized states $I_{\rm{est}} \sim 1/\lambda^2$ (see Ref.~\cite{schreiber2015}, supplementary material). The graph of $I_\infty$ vs. $\beta$ and $\Delta/J$ is fractal (see Fig.~S1 in the Supplementary Information), as it has different behaviors for rational and irrational $\beta$. Despite this complexity, the long time behavior of $I$ is distinct in the localized and delocalized phase: $I(t)$ captures the localization transition, but also probes features of the single-particle wave functions beyond the localization length.

The third term in Eq.~(\ref{hamileq}) describes on-site interactions of strength $U$.  Here we develop a low-density expansion to calculate the imbalance in the presence of interactions.

%Noting that the single particle states are exponentially localized, one can give the naive estimate $I(t \to \infty) \sim 1/\lambda^2 \sim \log^2 \frac{\Delta}{2 J} $ as $\lambda \to \infty$ \cite{schreiber2015}. This estimate seems to fail for large systems where we see instead that for some $\beta$, we have roughly $I(t \to \infty) \sim 1/\lambda$. Moreover, we observe strong $\beta$ dependence in the long-time imbalance; for instance, at $\Delta/J=3$, a plot of $I(t\to \infty)$ vs $\beta$ seems to produce a complex fractal structure (see Supplementary Information).%

%Recent theoretical work has shown that localization persists in the presence of interactions, even in the thermodynamic limit \cite{iyer2013}.  Our interest here however is not in the thermodynamic limit, but rather is in modeling the finite size systems of the experiment in Ref. \cite{schreiber2015}.%

%\subsection{Cluster Expansion Method} %

We define $\langle I(t)\rangle$ to be the expectation value of the imbalance, averaged over the ensemble of initial states,
\begin{equation} \label{eqgen}
\langle I(t)\rangle= \frac{1}{Z} \sum_{n=1}^{N_s} \sum_{\{n\}} W(\{n\}) \times \frac{1}{n} \langle \{n\}|  \hat{n}_I (t) | \{n\} \rangle
\end{equation}
Here $\{n\}= \{i_1 \sigma_1, i_2 \sigma_2, ..., i_n \sigma_n \}$ labels an $n$-particle initial state with particles at sites $i$ with spin $\sigma$, $\sum_{\{n\}}$ denotes a sum over the $i_j$'s and $\sigma_j$'s, $W(\{n\})$ is the weight of a given $n$ particle state, $Z= \sum_{\{n\}} W(\{n\})$, and $\hat{n}_I (t)= e^{i H t} (\hat{N}_{\textrm{odd}}- \hat{N}_{\textrm{even}}) e^{-i H t}$ where $\hat{N}_{\textrm{odd/even}} $ are the number operators (for both spins) on odd/even sites. 

\begin{figure}  \vspace{1.0em}
\hbox{\hspace{-1.4em}
\includegraphics[width=0.5\textwidth]{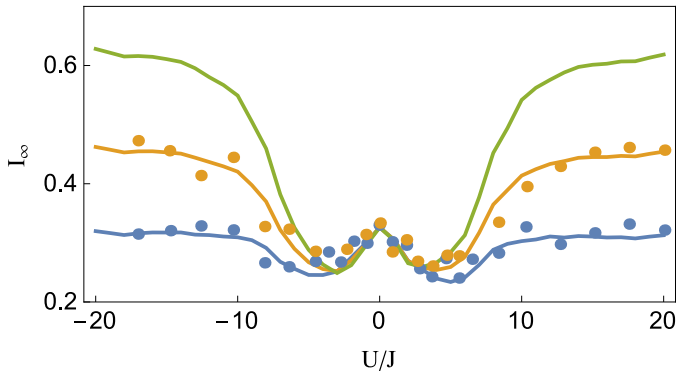}}
\caption{(Color online) Long time density imbalance $I_\infty$ as a function of interaction strength $U/J$ for a one-dimensional lattice at superlattice strength $\Delta/J= 3$. The different curves show calculations using a cluster expansion on a 20 site lattice with different densities  $\eta$ of doublons in the ensemble of initial states: The bottom (blue), middle (orange), and top (green) curves correspond to a ratio of doublons to particles of $\eta/\epsilon=0, 0.23, 0.5$, respectively. The blue and orange points are experimental measurements for a small doublon fraction ($\eta/\epsilon \approx 0.08$) and larger doublon fraction ($\eta/\epsilon \approx 0.5$), from Fig.~6 of Ref.~\cite{schreiber2015}, courtesy of Ulrich Schneider.} \label{doubfig} 
\end{figure}

To model the experiment, we take $W(\{n \}) =0$ if any of the particles are on even sites. We take the initial occupation of each odd site to be an independent random variable, and hence $W(\{n \})= \epsilon^n (1-\epsilon)^{N_s -n}$, where $N_s$ is the number of sites. Our method is readily generalized to more sophisticated weights. For instance, as shown in Eq.~(S12), we can weight the initial states with separate probabilities for sites with two atoms (doublons) or one atom (singlons) (see also Fig.~\ref{doubfig}).

% Here we choose the weighting function  $W(\{n\})= \epsilon^n (1-\epsilon)^{N_s- n}$, where $N_s$ is the number of sites. This weighting function corresponds to initializing the states by placing particles randomly on odd sites with probability $\epsilon$. We note that this choice of $W(\{n\})$ is primarily made for simplicity and that the calculations that follow can readily be generalized to other probability distributions. %

With this choice of $W$, the normalization is $Z= 1- (1-\epsilon)^{N_s}$ which approaches $1$ in the $N_s \to \infty$ limit. In that same limit, the mean density (the number of particles per site averaged over the ensemble of initial states) is $\epsilon$ .

Substituting our weight function into Eq.~(\ref{eqgen}) yields an expression for the imbalance as a sum of terms involving different numbers of particles:
\begin{widetext}
\begin{equation} \label{eqimb}
\begin{split}
& \langle I(t)\rangle=  \frac{1}{Z} \Big[  \epsilon(1-\epsilon)^{N_s-1} \sum_{\{1\}}{}^{'} C_{\{1\}}(t)  + \frac{\epsilon^2}{2}(1-\epsilon)^{N_s-2} \sum_{\{2\}} {}^{'} C_{\{2\}}(t) + \frac{\epsilon^3}{3}(1-\epsilon)^{N_s-3} \sum_{\{3\}} {}^{'} C_{\{3\}}(t)  + ... + \frac{\epsilon^{N_s}}{N_s}  \sum_{\{N_s\}} {}^{'} C_{\{ N_s \} } (t)  \Big]
\end{split}
\end{equation}
where $C_{\{n\}}(t)=  \langle \{n\}|  \hat{n}_I (t) | \{n\} \rangle$, and the primes on the sums mean they only include odd sites.

We wish to resum this series, taking advantage of the fact that well-separated particles will move independently. Somewhat analogous to cumulants, we define functions $\widetilde{C}_{\{ n \} } (t)$ via
\begin{equation} \label{tildeeq}
\begin{split}
& C_{\{n\}}(t) = \widetilde{C}_{\{ n\}}(t)+ \sum_{  \langle\{1\} \in \{n \} \rangle }{C}_{\{ 1\}}(t) + \sum_{  \langle\{2 \} \in \{n \} \rangle }\widetilde{C}_{\{ 2\}}(t)  + \sum_{  \langle\{3 \} \in \{n \} \rangle }\widetilde{C}_{\{ 3\}}(t)+ ... + \sum_{  \langle\{n-1 \} \in \{n \} \rangle }\widetilde{C}_{\{ n-1\}}(t)  \\
\end{split} 
\end{equation}
\end{widetext}
where $\sum_{  \langle\{k \} \in \{n \} \rangle }$ denotes a sum over all $\binom{n}{k}$ combinations of $k$ site and spin labels in $\{n\}$. We set $\widetilde{C}_{\{1\}} (t)= C_{\{1\}} (t)$. These new functions $\widetilde{C}_{\{k \} }(t)$ extract the $k$-body dynamics from the original functions $C_{ \{k\}}  (t)$. First instance, the two particle term $\widetilde{C}_{\{ i_1\sigma_1, i_2\sigma_2 \}} (t)=C_{\{ i_1\sigma_1, i_2\sigma_2 \}} (t) - C_{\{ i_1\sigma_1\}}(t) - C_{\{ i_2\sigma_2\}} (t)$ is the difference between a term representing the exact dynamics of two particles with initial positions and spins $i_1\sigma_1$ and $i_2\sigma_2$ and the single particle dynamics of a particle initialized at site $i_1$ and another particle initialized at site $i_2$. In the non-interacting limit $U=0$, we only have single particle dynamics and $\widetilde{C}_{\{k \}} (t)=0$ for all $k>1$. In a diagrammatic formulation, $\widetilde{C}$ involves only connected diagrams.

Substituting Eq.~(\ref{tildeeq}) into Eq.~(\ref{eqimb}), and using the arguments in the Supplementary Information gives
\begin{equation} \label{eqifinal}
\begin{split}
 \langle I(t)\rangle = \frac{1}{N_s} \sum_{\{1\}} {}^{'} \widetilde{C}_{\{ 1 \}}(t) +\frac{\epsilon}{N_s} \sum_{\{2 \}} {}^{'} \widetilde{C}_{\{ 2 \}}(t) + O(\epsilon^2)
\end{split}
\end{equation}
in the $N_s \to \infty$ limit. For our numerical calculations we include the finite size corrections in Eq.~(S7).

Equation~(\ref{eqifinal}) expresses the $n$-particle time dependent observable $\langle I(t)\rangle$ explicitly as the sum of 1-particle terms ($\widetilde{C}_{\{1\}}(t)$), 2-particle terms ($\widetilde{C}_{\{2\}}(t)$), etc. The first sum in Eq.~(\ref{eqifinal}) contains $N_s$ terms. The second sum contains $O(N_s^2)$ terms, but when the two particles are farther apart than some length scale $\xi$, where $\xi$ is the smaller of the one-particle localization length $\lambda$ and the ballistic length $l= J t$, the particles are effectively non-interacting and $\widetilde{C}_{\{2\}}$ will vanish. Therefore only $\xi N_s$ terms contribute to the sum. Similarly, there are only $O(\xi^2 N_s)$ which contribute in the sum over $\widetilde{C}_{\{3\}}$ terms. 

%However, the only terms which will contribute substantially to the sum are the $\approx \xi N_s$ terms where all three particles are within a length $\xi$ from each other. All other terms correspond to initial states of three widely spaced particles, or initial states with pairs of nearby particles which are at a distance greater than $\xi$ away from the third particle; for such terms, $\widetilde{C}_{\{3\}}$ is vanishingly small. %

Each subsequent term in Eq.~(\ref{eqifinal}) is intensive and is weighted by a coefficient of the order $\epsilon^{n-1}$ (the density exponentiated to the number of particles in the cluster minus 1). This cluster expansion is a non-equilibrium analogue to the virial expansion in statistical physics \cite{kardar2007}.  When the localization length is greater than the system size ($\lambda > N_s$) the series is only guaranteed to converge for short times $l=Jt \lesssim 1/\epsilon$. Therefore, for calculations of the long-time behavior of the imbalance, we focus our attention on the localized regime $\Delta/J >2$.

%The second contains $N_s^2$ terms, but $\widetilde{C}_{\{2\}}$ vanishes when the two particles start farther apart than some length scale $\xi$, where $\xi$ is the smaller of the single particle localization length $\lambda$ and the ballistic length $l= J t$. Therefore in the localized regime only $\lambda N_s$ terms are non-zero%

%Thus each contribution to Eq.~(\ref{eqifinal}) is intensive and is weighted by a coefficient of the order $\epsilon^{n-1}$ (the density exponentiated to the number of particles in the cluster minus 1), and for small $\epsilon$ the series is rapidly convergent.  This cluster expansion is a non-equilibrium analogue to the virial expansion in statistical physics \cite{kardar2007}.%

For most of the results in this paper we only keep the first two terms in Eq.~(\ref{eqifinal}). Remarkably, this approximation, which only involves calculating the dynamics of one or two particles, shows all the features seen in the experiments of Ref.~\cite{schreiber2015}.

\textit{Numerical Results} - Figure~\ref{modelfig} shows typical $\langle I(t) \rangle$ for interacting fermions in the localized regime. The solid blue curves show calculations using the first two terms in the cluster expansion in Eq.~(\ref{eqifinal}). The imbalance initially has a value $I(t=0)=1$, reflecting the fact that the initial states have particles localized only on odd sites. At long times, the imbalance saturates to a non-zero value with small fluctuations about the mean. For comparison, the red dots show calculations using time-dependent density matrix renormalization group (t-DMRG) \cite{white1993, white2004}. For the DMRG calculations, we average over 100 initial states drawn from the probability distribution $W(\{n \})$. The cluster expansion and the t-DMRG show excellent agreement at the smaller density $\epsilon=0.2$. At the larger density $\epsilon=0.5$ there is some small quantitative disagreement, but the average long-time imbalance is nearly identical for the two approaches. As a test of the convergence of the cluster expansion, we have also computed the contribution from three-particle terms (orange curve in Fig.~\ref{modelfig}). Including these terms gives small corrections to the two-particle calculation and yields better agreement with t-DMRG.

Figure~\ref{fig2} shows the long time imbalance $I_\infty$ as a function of interaction strength for a series of superlattice strengths. We compute $\langle I(t) \rangle_{\rm{init}}$ by numerically evaluating the first two terms of Eq.~(\ref{eqifinal}) at a density $\epsilon=0.2$. Each data point in Fig.~\ref{fig2} represents $\langle I(t) \rangle$ averaged over the times $200< tJ <500$ and averaged over twelve values of the superlattice potential phase $\phi$ evenly spaced in the range $[0, \pi]$. All simulations were performed on a lattice with 20 sites using open boundary conditions. We have explicitly verified that finite size effects are negligible; the system size was chosen for numerical convenience. 

Each curve is symmetric under $U \to -U$. As pointed out in Ref.~\cite{schneider2012} this symmetry is expected for time-reversal invariant operators such as $I(t)$, as long as the initial states are localized in space. For $|U/J| \lesssim 2\Delta$, interactions cause some 2-particle scattering states to become less localized than 1-particle states, and the long time imbalance decreases with increasing interaction strength. For larger interactions, the imbalance begins to increase again and produces a ``W" shape consistent with the re-entrant behavior predicted for similar systems \cite{michal2015}. The ``W" is most pronounced for $\Delta/J \approx 3$. 

At large interaction strengths, up-spin and down-spin particles initially localized at the same site (doublons) become bound and have a reduced effective tunneling rate $J_{\rm{eff}}\approx J^2/U$ \cite{barelli1996, dufour2012}. The contribution to $I_\infty$ from these doublons causes the long time imbalance at large interaction strengths to become greater than the long-time imbalance at $U=0$.

We further explore the contribution of doublons to $I_\infty$ by giving doublons and singlons separate weights in our average over initial states (see Eq.~(S12)). We let $\epsilon$ be the total density of particles and $\eta$ the density of doublons. Fig.~\ref{doubfig} shows $I_\infty$ as a function of $U/J$ at $\Delta/J=3$ for three different values of $\eta/\epsilon$ in the initial states of the system: $0$ ($\epsilon= 0.5$), $0.23$ ($\epsilon=0.57$), and $0.5$ ($\epsilon=0.67$) for the bottom (blue), middle (orange), top (green) graphs, respectively. All other parameters are the same as in Figure~\ref{fig2}. In the case where there are no doublons $I_\infty(U/J=0)= I_\infty(U/J \to \infty)$. This is a reflection of the fact that the dynamics of singlons in the hard core $U/J\to \infty$ limit is identical to the dynamics of free spinless fermions \cite{schreiber2015}. As more doublons are added to the system, $I_\infty$ at large $U/J$ increases, as expected from the reduced tunneling rate of bound pairs. The blue and orange points in Fig.~\ref{doubfig} show corresponding experimental results from Ref.~\cite{schreiber2015}, where the doublon density was controlled by varying the loading protocol.

We chose $\eta$ and $\epsilon$ to best match the experimental data, finding excellent agreement. Our best-fit value of $\eta$ is somewhat smaller than estimates in Ref.~\cite{schreiber2015}.  Similar discrepancies were seen in DMRG calculations \cite{schreiber2015}.

%Figure~\ref{fig3} shows the density dependence of $I$ at fixed disorder strength $\Delta/J=3.0$. All other parameters are the same as Fig.~\ref{fig2}. Using our cluster expansion the density is changed simply by varying the $\epsilon$ parameter in the coefficients in front of the $C_1$ and $\widetilde{C}_2$ terms Eq.~(\ref{eqifinal}). As the density increases, interactions play a larger role and the ``W" shape of the $I$ vs $U/J$ data becomes more pronounced. Our approximation of only keeping two-particle terms in the cluster expansion breaks down for larger densities.%

Motivated by more recent experiments \cite{bordia2015}, and as a further demonstration of our cluster method approach, we have extended our calculations to two-dimensional lattices. We consider a two-dimensional Hamiltonian with a one-dimensional superlattice potential $V=\Delta \cos(2\pi \beta i_x +\phi)$. As before, we take $J$ to be the hopping in the x-direction and $J_y$ the hopping in the y-direction. In this case we average over initial states where atoms are localized on odd sites in the x-direction and are in $k_y=0$ momentum eigenstates in the y-direction. This choice of initial states, which requires periodic boundary conditions in the y-direction, was chosen purely for numerical simplicity; we expect no qualitative changes if we initialize with spatially localized states and use open boundary conditions in the y-direction. We once again use Eq.~(\ref{eqifinal}) including only one-particle and two-particle terms to compute the even-odd imbalance in the x-direction.

\begin{figure}  \vspace{1.0em}
\hbox{\hspace{-1.4em}
\includegraphics[width=0.5\textwidth]{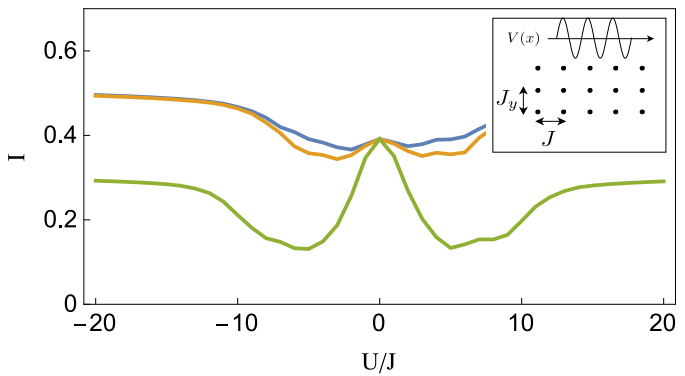}}
\caption{(Color online) Density imbalance $I$ averaged over time from $t=5/J$ to $t=10/J$ as a function of interaction strength $U/J$ for a two-dimensional lattice with 10$\times$10 sites at superlattice strength $\Delta/J= 3.0$ and density $\epsilon=0.2$. The superlattice potential is only one-dimensional: $V(i_x,i_y)= \Delta \cos(2\pi \beta i_x + \phi)$. $J_y/J=0, 0.1, 1$ for the top, middle, and bottom (blue, orange, green) curves, respectively. The inset shows a diagram of the setup.} \label{twodfig}
\end{figure}

Because the eigenstates are inherently delocalized in this situation, we only expect our cluster expansion to be accurate for short times. Fig.~\ref{twodfig} shows the imbalance $I$ in the x-direction, averaged over times between $t=5/J$ and $t=10/J$ as a function of $U/J$. These simulations were performed on a lattice with 10$\times$10 sites. Scattering in the y-direction (transverse to the superlattice potential) allows for the density imbalance to relax to smaller values, and $I$ becomes suppressed as $J_y$ is increased. Similar results are observed in Ref.~\cite{bordia2015}.

% \begin{figure} \vspace{1.0em}
%\hbox{\hspace{-1.4em}
%\includegraphics[width=0.5\textwidth]{mblocal_zoomfig.eps}}
%\caption{(Color online) Long time density imbalance $I_\infty$ as a function of interaction strength $U/J$ for a one-dimensional lattice with 20 sites at disorder strength $\Delta/J= 3.0$. This plot %shows a characteristic ``W" shape consistent with experimental observations \cite{schreiber2015}.}  \label{fig3}
%\end{figure}

\textit{Conclusion} - In this paper we have applied a new cluster expansion method to simulate experiments \cite{schreiber2015} which studied the non-equilibrium dynamics of fermions pattern-loaded in quasi-disordered one-dimensional lattices. Our calculations, which involve keeping the first two terms in the cluster expansion and account for only single particle and two particle dynamics, reproduce all experimental features of the long-time density imbalance between even and odd sites, and agree quantitatively with simulations using t-DMRG. We have also extended our calculations to two-dimensional lattices, finding that the density imbalance is suppressed when adding hopping in the direction transverse to the superlattice potential.

Although principally designed to calculate the experimental observable, this cluster approach also gives some insight into many-body localization. For example we have shown that time dynamics of the many-body wave function in the localized phase can be written as a sum of 1-body, 2-body, ..., n-body terms. In the dilute limit, the dynamics are dominated by few-particle physics, a feature which was not previously recognized.

Our cluster approach can be also used to explicitly construct the local integrals of motion which underly the phenomenology of the many-body localized phase \cite{serbyn2013, huse2014, chandran2015, ros2015}.
As detailed below, we use the solution to the $j$-body problem to construct fermionic creation operators $a^{\dagger (j)}_{n \sigma}$ where $\{a^{ (j)}_{n \sigma}, a^{\dagger (j)}_{m \tau} \} = \delta_{m n} \delta_{\tau \sigma}$. Our operators have the property that in the $i$-particle subspace, all of the $a^{\dagger (j)}_{n \sigma}$ are equivalent for $j \geq i$: $a^{\dagger (i)}_{n \sigma} P_i = a^{\dagger (j)}_{n \sigma} P_i$ where $P_i$ projects into the $i$ particle subspace. Our conserved quantities are manifest in the requirement 
\begin{equation} \label{conseq}
[a^{ \dagger (i)}_{n \sigma} a^{ (i)}_{n \sigma}, P_i H P_i ] =0
\end{equation}
If the $a^{ \dagger (i)}_{n \sigma}$ are ``local", we thereby complete the construction.

We take $a^{ \dagger (1)}_{n \sigma}$ to create the single-particle eigenstate with spin $\sigma$ and energy $\epsilon_n$; suppressing the spin indices $|n \rangle = a^{ \dagger (1)}_{n } |\rm{vac} \rangle$. This operator is local if these eigenstates are localized. Trivially, Eq.~(\ref{conseq}) is satisfied. 

Next we construct 
\begin{equation}
 a^{\dagger (2)}_{n \sigma} =  a^{\dagger (1)}_{n \sigma} + \sum{\substack{jkl\\ \tau \tau' \tau''}} \Gamma^{n \sigma}_{\substack{jkl\\ \tau \tau' \tau''}} a^{\dagger (1)}_{j \tau} a^{\dagger (1)}_{k \tau'} a^{(1)}_{n \tau''}
\end{equation}
so that $ a^{\dagger (2)}_{n \sigma} P_1 = a^{\dagger (1)}_{n \sigma} P_1$. We can always choose the $\Gamma$'s such that $|n \sigma, m \tau \rangle = a^{\dagger (2)}_{n \sigma} a^{\dagger (2)}_{m \tau} | \rm{vac} \rangle$ is an eigenstate of $H$ with energy $E^{\sigma \tau}_{m n}$. Neglecting the spin indices
\begin{equation}
\Gamma^{n}_{j k l}= (\langle j| \otimes \langle k|) | n l \rangle - \delta_{j n} \delta_{k l}
\end{equation}

There are as many ways of doing this are there are ways of assigning the indices to the 2-particle states. We choose the indices to maximize the overlap $(\langle n| \otimes \langle m|) | n l \rangle$. If the two-particle states and one-particle states are localized, then $a^{\dagger (2)}_{n \sigma}$ will be localized. Eq.~(\ref{conseq}) is clearly satisfied. Constructing the higher order operators follows the same procedure.

To connect with the existing literature \cite{serbyn2013, huse2014, chandran2015, ros2015}, we note that this construction yields a Hamiltonian of the form 
\begin{equation} 
H= \sum_{n\sigma} \epsilon_n \tilde{n}_{n\sigma} + \sum_{\substack{nm\\ \sigma \sigma'}} U^{(2)}_{\substack{nm\\ \sigma \sigma'}} \tilde{n}_{n\sigma}  \tilde{n}_{m\sigma'}  + ... 
\end{equation}
where $\tilde{n}_{n \sigma} = \lim_{j\to\infty} a^{ \dagger (j)}_{n \sigma} a^{ (j)}_{n \sigma}$. The coefficients are local, meaning $U^{(k)}_{i_1 i_2...i_k} \sim \exp{(- \rm{max} |i_{\alpha}-i_{\beta}|/ \xi_k )}$. They can be expressed in terms of the eigenvalues of the $k$-body problem; for example $ U^{(2)}_{\substack{nm\\ \sigma \sigma' }}= E^{\sigma \sigma'}_{m n}- \epsilon_n - \epsilon_m$. The Supplementary Information shows a graph of this quantity for typical parameters, illustrating the exponential decay.

\textit{Acknowledgements}- We acknowledge support from ARO-MURI Non-equilibrium Many-body Dynamics grant (W911NF-14-1-0003). We thank Mark Fischer for discussions, and Ulrich Schneider for
sharing the experimental data.

\bibliography{mblocal}

\begin{thebibliography}{36}
\expandafter\ifx\csname natexlab\endcsname\relax\def\natexlab#1{#1}\fi
\expandafter\ifx\csname bibnamefont\endcsname\relax
  \def\bibnamefont#1{#1}\fi
\expandafter\ifx\csname bibfnamefont\endcsname\relax
  \def\bibfnamefont#1{#1}\fi
\expandafter\ifx\csname citenamefont\endcsname\relax
  \def\citenamefont#1{#1}\fi
\expandafter\ifx\csname url\endcsname\relax
  \def\url#1{\texttt{#1}}\fi
\expandafter\ifx\csname urlprefix\endcsname\relax\def\urlprefix{URL }\fi
\providecommand{\bibinfo}[2]{#2}
\providecommand{\eprint}[2][]{\url{#2}}

\bibitem[{\citenamefont{Anderson}(1958)}]{anderson1958}
\bibinfo{author}{\bibfnamefont{P.~W.} \bibnamefont{Anderson}},
  \bibinfo{journal}{Physical review} \textbf{\bibinfo{volume}{109}},
  \bibinfo{pages}{1492} (\bibinfo{year}{1958}).

\bibitem[{\citenamefont{Abrahams et~al.}(1979)\citenamefont{Abrahams, Anderson,
  Licciardello, and Ramakrishnan}}]{abrahams1979}
\bibinfo{author}{\bibfnamefont{E.}~\bibnamefont{Abrahams}},
  \bibinfo{author}{\bibfnamefont{P.}~\bibnamefont{Anderson}},
  \bibinfo{author}{\bibfnamefont{D.}~\bibnamefont{Licciardello}},
  \bibnamefont{and}
  \bibinfo{author}{\bibfnamefont{T.}~\bibnamefont{Ramakrishnan}},
  \bibinfo{journal}{Physical Review Letters} \textbf{\bibinfo{volume}{42}},
  \bibinfo{pages}{673} (\bibinfo{year}{1979}).

\bibitem[{\citenamefont{Billy et~al.}(2008)\citenamefont{Billy, Josse, Zuo,
  Bernard, Hambrecht, Lugan, Cl{\'e}ment, Sanchez-Palencia, Bouyer, and
  Aspect}}]{billy2008}
\bibinfo{author}{\bibfnamefont{J.}~\bibnamefont{Billy}},
  \bibinfo{author}{\bibfnamefont{V.}~\bibnamefont{Josse}},
  \bibinfo{author}{\bibfnamefont{Z.}~\bibnamefont{Zuo}},
  \bibinfo{author}{\bibfnamefont{A.}~\bibnamefont{Bernard}},
  \bibinfo{author}{\bibfnamefont{B.}~\bibnamefont{Hambrecht}},
  \bibinfo{author}{\bibfnamefont{P.}~\bibnamefont{Lugan}},
  \bibinfo{author}{\bibfnamefont{D.}~\bibnamefont{Cl{\'e}ment}},
  \bibinfo{author}{\bibfnamefont{L.}~\bibnamefont{Sanchez-Palencia}},
  \bibinfo{author}{\bibfnamefont{P.}~\bibnamefont{Bouyer}}, \bibnamefont{and}
  \bibinfo{author}{\bibfnamefont{A.}~\bibnamefont{Aspect}},
  \bibinfo{journal}{Nature} \textbf{\bibinfo{volume}{453}},
  \bibinfo{pages}{891} (\bibinfo{year}{2008}).

\bibitem[{\citenamefont{Kondov et~al.}(2011)\citenamefont{Kondov, McGehee,
  Zirbel, and DeMarco}}]{kondov2011}
\bibinfo{author}{\bibfnamefont{S.}~\bibnamefont{Kondov}},
  \bibinfo{author}{\bibfnamefont{W.}~\bibnamefont{McGehee}},
  \bibinfo{author}{\bibfnamefont{J.}~\bibnamefont{Zirbel}}, \bibnamefont{and}
  \bibinfo{author}{\bibfnamefont{B.}~\bibnamefont{DeMarco}},
  \bibinfo{journal}{Science} \textbf{\bibinfo{volume}{334}},
  \bibinfo{pages}{66} (\bibinfo{year}{2011}).

\bibitem[{\citenamefont{Jendrzejewski et~al.}(2012)\citenamefont{Jendrzejewski,
  Bernard, Mueller, Cheinet, Josse, Piraud, Pezz{\'e}, Sanchez-Palencia,
  Aspect, and Bouyer}}]{jendrzejewski2012}
\bibinfo{author}{\bibfnamefont{F.}~\bibnamefont{Jendrzejewski}},
  \bibinfo{author}{\bibfnamefont{A.}~\bibnamefont{Bernard}},
  \bibinfo{author}{\bibfnamefont{K.}~\bibnamefont{Mueller}},
  \bibinfo{author}{\bibfnamefont{P.}~\bibnamefont{Cheinet}},
  \bibinfo{author}{\bibfnamefont{V.}~\bibnamefont{Josse}},
  \bibinfo{author}{\bibfnamefont{M.}~\bibnamefont{Piraud}},
  \bibinfo{author}{\bibfnamefont{L.}~\bibnamefont{Pezz{\'e}}},
  \bibinfo{author}{\bibfnamefont{L.}~\bibnamefont{Sanchez-Palencia}},
  \bibinfo{author}{\bibfnamefont{A.}~\bibnamefont{Aspect}}, \bibnamefont{and}
  \bibinfo{author}{\bibfnamefont{P.}~\bibnamefont{Bouyer}},
  \bibinfo{journal}{Nature Physics} \textbf{\bibinfo{volume}{8}},
  \bibinfo{pages}{398} (\bibinfo{year}{2012}).

\bibitem[{\citenamefont{Roati et~al.}(2008)\citenamefont{Roati, DÕErrico,
  Fallani, Fattori, Fort, Zaccanti, Modugno, Modugno, and
  Inguscio}}]{roati2008}
\bibinfo{author}{\bibfnamefont{G.}~\bibnamefont{Roati}},
  \bibinfo{author}{\bibfnamefont{C.}~\bibnamefont{DÕErrico}},
  \bibinfo{author}{\bibfnamefont{L.}~\bibnamefont{Fallani}},
  \bibinfo{author}{\bibfnamefont{M.}~\bibnamefont{Fattori}},
  \bibinfo{author}{\bibfnamefont{C.}~\bibnamefont{Fort}},
  \bibinfo{author}{\bibfnamefont{M.}~\bibnamefont{Zaccanti}},
  \bibinfo{author}{\bibfnamefont{G.}~\bibnamefont{Modugno}},
  \bibinfo{author}{\bibfnamefont{M.}~\bibnamefont{Modugno}}, \bibnamefont{and}
  \bibinfo{author}{\bibfnamefont{M.}~\bibnamefont{Inguscio}},
  \bibinfo{journal}{Nature} \textbf{\bibinfo{volume}{453}},
  \bibinfo{pages}{895} (\bibinfo{year}{2008}).

\bibitem[{\citenamefont{Shepelyansky}(1996)}]{shepelyansky1996}
\bibinfo{author}{\bibfnamefont{D.}~\bibnamefont{Shepelyansky}},
  \bibinfo{journal}{Physical Review B} \textbf{\bibinfo{volume}{54}},
  \bibinfo{pages}{14896} (\bibinfo{year}{1996}).

\bibitem[{\citenamefont{Barelli et~al.}(1996)\citenamefont{Barelli, Bellissard,
  Jacquod, and Shepelyansky}}]{barelli1996}
\bibinfo{author}{\bibfnamefont{A.}~\bibnamefont{Barelli}},
  \bibinfo{author}{\bibfnamefont{J.}~\bibnamefont{Bellissard}},
  \bibinfo{author}{\bibfnamefont{P.}~\bibnamefont{Jacquod}}, \bibnamefont{and}
  \bibinfo{author}{\bibfnamefont{D.~L.} \bibnamefont{Shepelyansky}},
  \bibinfo{journal}{Physical review letters} \textbf{\bibinfo{volume}{77}},
  \bibinfo{pages}{4752} (\bibinfo{year}{1996}).

\bibitem[{\citenamefont{Eilmes et~al.}(1999)\citenamefont{Eilmes, Grimm,
  R{\"o}mer, and Schreiber}}]{eilmes1999}
\bibinfo{author}{\bibfnamefont{A.}~\bibnamefont{Eilmes}},
  \bibinfo{author}{\bibfnamefont{U.}~\bibnamefont{Grimm}},
  \bibinfo{author}{\bibfnamefont{R.~A.} \bibnamefont{R{\"o}mer}},
  \bibnamefont{and}
  \bibinfo{author}{\bibfnamefont{M.}~\bibnamefont{Schreiber}},
  \bibinfo{journal}{The European Physical Journal B-Condensed Matter and
  Complex Systems} \textbf{\bibinfo{volume}{8}}, \bibinfo{pages}{547}
  (\bibinfo{year}{1999}).

\bibitem[{\citenamefont{Gornyi et~al.}(2005)\citenamefont{Gornyi, Mirlin, and
  Polyakov}}]{gornyi2005}
\bibinfo{author}{\bibfnamefont{I.}~\bibnamefont{Gornyi}},
  \bibinfo{author}{\bibfnamefont{A.}~\bibnamefont{Mirlin}}, \bibnamefont{and}
  \bibinfo{author}{\bibfnamefont{D.}~\bibnamefont{Polyakov}},
  \bibinfo{journal}{Physical review letters} \textbf{\bibinfo{volume}{95}},
  \bibinfo{pages}{206603} (\bibinfo{year}{2005}).

\bibitem[{\citenamefont{Basko et~al.}(2006)\citenamefont{Basko, Aleiner, and
  Altshuler}}]{basko2006}
\bibinfo{author}{\bibfnamefont{D.}~\bibnamefont{Basko}},
  \bibinfo{author}{\bibfnamefont{I.}~\bibnamefont{Aleiner}}, \bibnamefont{and}
  \bibinfo{author}{\bibfnamefont{B.}~\bibnamefont{Altshuler}},
  \bibinfo{journal}{Annals of physics} \textbf{\bibinfo{volume}{321}},
  \bibinfo{pages}{1126} (\bibinfo{year}{2006}).

\bibitem[{\citenamefont{Dufour and Orso}(2012)}]{dufour2012}
\bibinfo{author}{\bibfnamefont{G.}~\bibnamefont{Dufour}} \bibnamefont{and}
  \bibinfo{author}{\bibfnamefont{G.}~\bibnamefont{Orso}},
  \bibinfo{journal}{Physical review letters} \textbf{\bibinfo{volume}{109}},
  \bibinfo{pages}{155306} (\bibinfo{year}{2012}).

\bibitem[{\citenamefont{Tezuka and Garc{\'\i}a-Garc{\'\i}a}(2012)}]{tezuka2012}
\bibinfo{author}{\bibfnamefont{M.}~\bibnamefont{Tezuka}} \bibnamefont{and}
  \bibinfo{author}{\bibfnamefont{A.~M.} \bibnamefont{Garc{\'\i}a-Garc{\'\i}a}},
  \bibinfo{journal}{Physical Review A} \textbf{\bibinfo{volume}{85}},
  \bibinfo{pages}{031602} (\bibinfo{year}{2012}).

\bibitem[{\citenamefont{Iyer et~al.}(2013)\citenamefont{Iyer, Oganesyan,
  Refael, and Huse}}]{iyer2013}
\bibinfo{author}{\bibfnamefont{S.}~\bibnamefont{Iyer}},
  \bibinfo{author}{\bibfnamefont{V.}~\bibnamefont{Oganesyan}},
  \bibinfo{author}{\bibfnamefont{G.}~\bibnamefont{Refael}}, \bibnamefont{and}
  \bibinfo{author}{\bibfnamefont{D.~A.} \bibnamefont{Huse}},
  \bibinfo{journal}{Physical Review B} \textbf{\bibinfo{volume}{87}},
  \bibinfo{pages}{134202} (\bibinfo{year}{2013}).

\bibitem[{\citenamefont{Serbyn et~al.}(2013)\citenamefont{Serbyn, Papi{\'c},
  and Abanin}}]{serbyn2013}
\bibinfo{author}{\bibfnamefont{M.}~\bibnamefont{Serbyn}},
  \bibinfo{author}{\bibfnamefont{Z.}~\bibnamefont{Papi{\'c}}},
  \bibnamefont{and} \bibinfo{author}{\bibfnamefont{D.~A.}
  \bibnamefont{Abanin}}, \bibinfo{journal}{Physical review letters}
  \textbf{\bibinfo{volume}{111}}, \bibinfo{pages}{127201}
  (\bibinfo{year}{2013}).

\bibitem[{\citenamefont{Serbyn et~al.}(2014)\citenamefont{Serbyn, Papi{\'c},
  and Abanin}}]{serbyn2014}
\bibinfo{author}{\bibfnamefont{M.}~\bibnamefont{Serbyn}},
  \bibinfo{author}{\bibfnamefont{Z.}~\bibnamefont{Papi{\'c}}},
  \bibnamefont{and} \bibinfo{author}{\bibfnamefont{D.~A.}
  \bibnamefont{Abanin}}, \bibinfo{journal}{Physical Review B}
  \textbf{\bibinfo{volume}{90}}, \bibinfo{pages}{174302}
  (\bibinfo{year}{2014}).

\bibitem[{\citenamefont{Huse et~al.}(2014)\citenamefont{Huse, Nandkishore, and
  Oganesyan}}]{huse2014}
\bibinfo{author}{\bibfnamefont{D.~A.} \bibnamefont{Huse}},
  \bibinfo{author}{\bibfnamefont{R.}~\bibnamefont{Nandkishore}},
  \bibnamefont{and}
  \bibinfo{author}{\bibfnamefont{V.}~\bibnamefont{Oganesyan}},
  \bibinfo{journal}{Physical Review B} \textbf{\bibinfo{volume}{90}},
  \bibinfo{pages}{174202} (\bibinfo{year}{2014}).

\bibitem[{\citenamefont{Vosk et~al.}(2014)\citenamefont{Vosk, Huse, and
  Altman}}]{vosk2014}
\bibinfo{author}{\bibfnamefont{R.}~\bibnamefont{Vosk}},
  \bibinfo{author}{\bibfnamefont{D.~A.} \bibnamefont{Huse}}, \bibnamefont{and}
  \bibinfo{author}{\bibfnamefont{E.}~\bibnamefont{Altman}},
  \bibinfo{journal}{arXiv preprint arXiv:1412.3117}  (\bibinfo{year}{2014}).

\bibitem[{\citenamefont{Altman and Vosk}(2015)}]{altman2015}
\bibinfo{author}{\bibfnamefont{E.}~\bibnamefont{Altman}} \bibnamefont{and}
  \bibinfo{author}{\bibfnamefont{R.}~\bibnamefont{Vosk}},
  \bibinfo{journal}{Annu. Rev. Condens. Matter Phys.}
  \textbf{\bibinfo{volume}{6}}, \bibinfo{pages}{383} (\bibinfo{year}{2015}).

\bibitem[{\citenamefont{Li et~al.}(2015)\citenamefont{Li, Ganeshan, Pixley, and
  Sarma}}]{li2015}
\bibinfo{author}{\bibfnamefont{X.}~\bibnamefont{Li}},
  \bibinfo{author}{\bibfnamefont{S.}~\bibnamefont{Ganeshan}},
  \bibinfo{author}{\bibfnamefont{J.}~\bibnamefont{Pixley}}, \bibnamefont{and}
  \bibinfo{author}{\bibfnamefont{S.~D.} \bibnamefont{Sarma}},
  \bibinfo{journal}{arXiv preprint arXiv:1504.00016}  (\bibinfo{year}{2015}).

\bibitem[{\citenamefont{Modak and Mukerjee}(2015)}]{modak2015}
\bibinfo{author}{\bibfnamefont{R.}~\bibnamefont{Modak}} \bibnamefont{and}
  \bibinfo{author}{\bibfnamefont{S.}~\bibnamefont{Mukerjee}},
  \bibinfo{journal}{arXiv preprint arXiv:1503.07620}  (\bibinfo{year}{2015}).

\bibitem[{\citenamefont{Wang et~al.}(2015)\citenamefont{Wang, Hu, and
  Chen}}]{wang2015}
\bibinfo{author}{\bibfnamefont{Y.}~\bibnamefont{Wang}},
  \bibinfo{author}{\bibfnamefont{H.}~\bibnamefont{Hu}}, \bibnamefont{and}
  \bibinfo{author}{\bibfnamefont{S.}~\bibnamefont{Chen}},
  \bibinfo{journal}{arXiv preprint arXiv:1505.06343}  (\bibinfo{year}{2015}).

\bibitem[{\citenamefont{Nandkishore and Huse}(2014)}]{nandkishore2014}
\bibinfo{author}{\bibfnamefont{R.}~\bibnamefont{Nandkishore}} \bibnamefont{and}
  \bibinfo{author}{\bibfnamefont{D.~A.} \bibnamefont{Huse}},
  \bibinfo{journal}{Annual Review of Condensed Matter Physics}
  \textbf{\bibinfo{volume}{6}}, \bibinfo{pages}{15} (\bibinfo{year}{2014}).

\bibitem[{\citenamefont{Eisert et~al.}(2015)\citenamefont{Eisert, Friesdorf,
  and Gogolin}}]{eisert2015}
\bibinfo{author}{\bibfnamefont{J.}~\bibnamefont{Eisert}},
  \bibinfo{author}{\bibfnamefont{M.}~\bibnamefont{Friesdorf}},
  \bibnamefont{and} \bibinfo{author}{\bibfnamefont{C.}~\bibnamefont{Gogolin}},
  \bibinfo{journal}{Nature Physics} \textbf{\bibinfo{volume}{11}},
  \bibinfo{pages}{124} (\bibinfo{year}{2015}).

\bibitem[{\citenamefont{Devakul and Singh}(2015)}]{devakul2015}
\bibinfo{author}{\bibfnamefont{T.}~\bibnamefont{Devakul}} \bibnamefont{and}
  \bibinfo{author}{\bibfnamefont{R.~R.} \bibnamefont{Singh}},
  \bibinfo{journal}{Physical Review Letters} \textbf{\bibinfo{volume}{115}},
  \bibinfo{pages}{187201} (\bibinfo{year}{2015}).

\bibitem[{\citenamefont{Schreiber et~al.}(2015)\citenamefont{Schreiber,
  Hodgman, Bordia, LŸschen, Fischer, Vosk, Altman, Schneider, and
  Bloch}}]{schreiber2015}
\bibinfo{author}{\bibfnamefont{M.}~\bibnamefont{Schreiber}},
  \bibinfo{author}{\bibfnamefont{S.~S.} \bibnamefont{Hodgman}},
  \bibinfo{author}{\bibfnamefont{P.}~\bibnamefont{Bordia}},
  \bibinfo{author}{\bibfnamefont{H.~P.} \bibnamefont{LŸschen}},
  \bibinfo{author}{\bibfnamefont{M.~H.} \bibnamefont{Fischer}},
  \bibinfo{author}{\bibfnamefont{R.}~\bibnamefont{Vosk}},
  \bibinfo{author}{\bibfnamefont{E.}~\bibnamefont{Altman}},
  \bibinfo{author}{\bibfnamefont{U.}~\bibnamefont{Schneider}},
  \bibnamefont{and} \bibinfo{author}{\bibfnamefont{I.}~\bibnamefont{Bloch}},
  \bibinfo{journal}{Science} \textbf{\bibinfo{volume}{349}},
  \bibinfo{pages}{842} (\bibinfo{year}{2015}).

\bibitem[{\citenamefont{Aubry and Andr{\'e}}(1980)}]{aubry1980}
\bibinfo{author}{\bibfnamefont{S.}~\bibnamefont{Aubry}} \bibnamefont{and}
  \bibinfo{author}{\bibfnamefont{G.}~\bibnamefont{Andr{\'e}}},
  \bibinfo{journal}{Ann. Israel Phys. Soc} \textbf{\bibinfo{volume}{3}},
  \bibinfo{pages}{18} (\bibinfo{year}{1980}).

\bibitem[{\citenamefont{Sokoloff}(1985)}]{sokoloff1985}
\bibinfo{author}{\bibfnamefont{J.}~\bibnamefont{Sokoloff}},
  \bibinfo{journal}{Physics Reports} \textbf{\bibinfo{volume}{126}},
  \bibinfo{pages}{189} (\bibinfo{year}{1985}).

\bibitem[{\citenamefont{Kardar}(2007)}]{kardar2007}
\bibinfo{author}{\bibfnamefont{M.}~\bibnamefont{Kardar}},
  \emph{\bibinfo{title}{Statistical physics of particles}}
  (\bibinfo{publisher}{Cambridge University Press}, \bibinfo{year}{2007}).

\bibitem[{\citenamefont{White}(1993)}]{white1993}
\bibinfo{author}{\bibfnamefont{S.~R.} \bibnamefont{White}},
  \bibinfo{journal}{Physical Review B} \textbf{\bibinfo{volume}{48}},
  \bibinfo{pages}{10345} (\bibinfo{year}{1993}).

\bibitem[{\citenamefont{White and Feiguin}(2004)}]{white2004}
\bibinfo{author}{\bibfnamefont{S.~R.} \bibnamefont{White}} \bibnamefont{and}
  \bibinfo{author}{\bibfnamefont{A.~E.} \bibnamefont{Feiguin}},
  \bibinfo{journal}{Physical review letters} \textbf{\bibinfo{volume}{93}},
  \bibinfo{pages}{076401} (\bibinfo{year}{2004}).

\bibitem[{\citenamefont{Schneider et~al.}(2012)\citenamefont{Schneider,
  Hackerm{\"u}ller, Ronzheimer, Will, Braun, Best, Bloch, Demler, Mandt, Rasch
  et~al.}}]{schneider2012}
\bibinfo{author}{\bibfnamefont{U.}~\bibnamefont{Schneider}},
  \bibinfo{author}{\bibfnamefont{L.}~\bibnamefont{Hackerm{\"u}ller}},
  \bibinfo{author}{\bibfnamefont{J.~P.} \bibnamefont{Ronzheimer}},
  \bibinfo{author}{\bibfnamefont{S.}~\bibnamefont{Will}},
  \bibinfo{author}{\bibfnamefont{S.}~\bibnamefont{Braun}},
  \bibinfo{author}{\bibfnamefont{T.}~\bibnamefont{Best}},
  \bibinfo{author}{\bibfnamefont{I.}~\bibnamefont{Bloch}},
  \bibinfo{author}{\bibfnamefont{E.}~\bibnamefont{Demler}},
  \bibinfo{author}{\bibfnamefont{S.}~\bibnamefont{Mandt}},
  \bibinfo{author}{\bibfnamefont{D.}~\bibnamefont{Rasch}},
  \bibnamefont{et~al.}, \bibinfo{journal}{Nature Physics}
  \textbf{\bibinfo{volume}{8}}, \bibinfo{pages}{213} (\bibinfo{year}{2012}).

\bibitem[{\citenamefont{Michal et~al.}(2015)\citenamefont{Michal, Aleiner,
  Altshuler, and Shlyapnikov}}]{michal2015}
\bibinfo{author}{\bibfnamefont{V.}~\bibnamefont{Michal}},
  \bibinfo{author}{\bibfnamefont{I.}~\bibnamefont{Aleiner}},
  \bibinfo{author}{\bibfnamefont{B.}~\bibnamefont{Altshuler}},
  \bibnamefont{and}
  \bibinfo{author}{\bibfnamefont{G.}~\bibnamefont{Shlyapnikov}},
  \bibinfo{journal}{arXiv preprint arXiv:1502.00282}  (\bibinfo{year}{2015}).

\bibitem[{\citenamefont{Bordia et~al.}(2015)\citenamefont{Bordia, L{\"u}schen,
  Hodgman, Schreiber, Bloch, and Schneider}}]{bordia2015}
\bibinfo{author}{\bibfnamefont{P.}~\bibnamefont{Bordia}},
  \bibinfo{author}{\bibfnamefont{H.~P.} \bibnamefont{L{\"u}schen}},
  \bibinfo{author}{\bibfnamefont{S.~S.} \bibnamefont{Hodgman}},
  \bibinfo{author}{\bibfnamefont{M.}~\bibnamefont{Schreiber}},
  \bibinfo{author}{\bibfnamefont{I.}~\bibnamefont{Bloch}}, \bibnamefont{and}
  \bibinfo{author}{\bibfnamefont{U.}~\bibnamefont{Schneider}},
  \bibinfo{journal}{arXiv preprint arXiv:1509.00478}  (\bibinfo{year}{2015}).

\bibitem[{\citenamefont{Chandran et~al.}(2015)\citenamefont{Chandran, Kim,
  Vidal, and Abanin}}]{chandran2015}
\bibinfo{author}{\bibfnamefont{A.}~\bibnamefont{Chandran}},
  \bibinfo{author}{\bibfnamefont{I.~H.} \bibnamefont{Kim}},
  \bibinfo{author}{\bibfnamefont{G.}~\bibnamefont{Vidal}}, \bibnamefont{and}
  \bibinfo{author}{\bibfnamefont{D.~A.} \bibnamefont{Abanin}},
  \bibinfo{journal}{Physical Review B} \textbf{\bibinfo{volume}{91}},
  \bibinfo{pages}{085425} (\bibinfo{year}{2015}).

\bibitem[{\citenamefont{Ros et~al.}(2015)\citenamefont{Ros, Mueller, and
  Scardicchio}}]{ros2015}
\bibinfo{author}{\bibfnamefont{V.}~\bibnamefont{Ros}},
  \bibinfo{author}{\bibfnamefont{M.}~\bibnamefont{Mueller}}, \bibnamefont{and}
  \bibinfo{author}{\bibfnamefont{A.}~\bibnamefont{Scardicchio}},
  \bibinfo{journal}{Nuclear Physics B} \textbf{\bibinfo{volume}{891}},
  \bibinfo{pages}{420} (\bibinfo{year}{2015}).

\end{thebibliography}

\onecolumngrid

\setcounter{equation}{0}
\setcounter{figure}{0}
\setcounter{table}{0}

\renewcommand{\theequation}{S\arabic{equation}}
\renewcommand{\thefigure}{S\arabic{figure}}

\section{Supplementary Information}

\subsection{Imbalance vs. Superlattice Period in the Non-interacting Limit}
In the non-interacting limit, the experiment is well modeled by the Aubrey-Andre model
\begin{equation}
H= -J \sum_{i, \sigma} \left(c^\dagger_{i, \sigma} c_{i+1, \sigma} + \mbox{h.c} \right)   + \Delta \sum_{i, \sigma}  \cos(2\pi \beta i +\phi) c^\dagger_{i, \sigma} c_{i, \sigma} 
\end{equation}
where $J$ is the nearest neighbor hopping strength, $\Delta$ is the strength of the periodic superlattice, and $\beta^{-1}$ is the period of the superlattice. As discussed in the main text, this is an interesting model as its behavior depends on if $\beta$ is rational or irrational (or in a finite system of length $N_s$, if $N_s\beta$ is an integer or not).

Starting with a particle on an odd site, we numerical evolve the single-particle wave-function and calculate the average long-time imbalance $I_\infty= n_{\textrm{odd}}- n_{\textrm{even}}$, where $n_{\textrm{odd/even}}$ is the average long-time density on odd and even sites, respectively.

\begin{figure}
\includegraphics[width=0.5\textwidth]{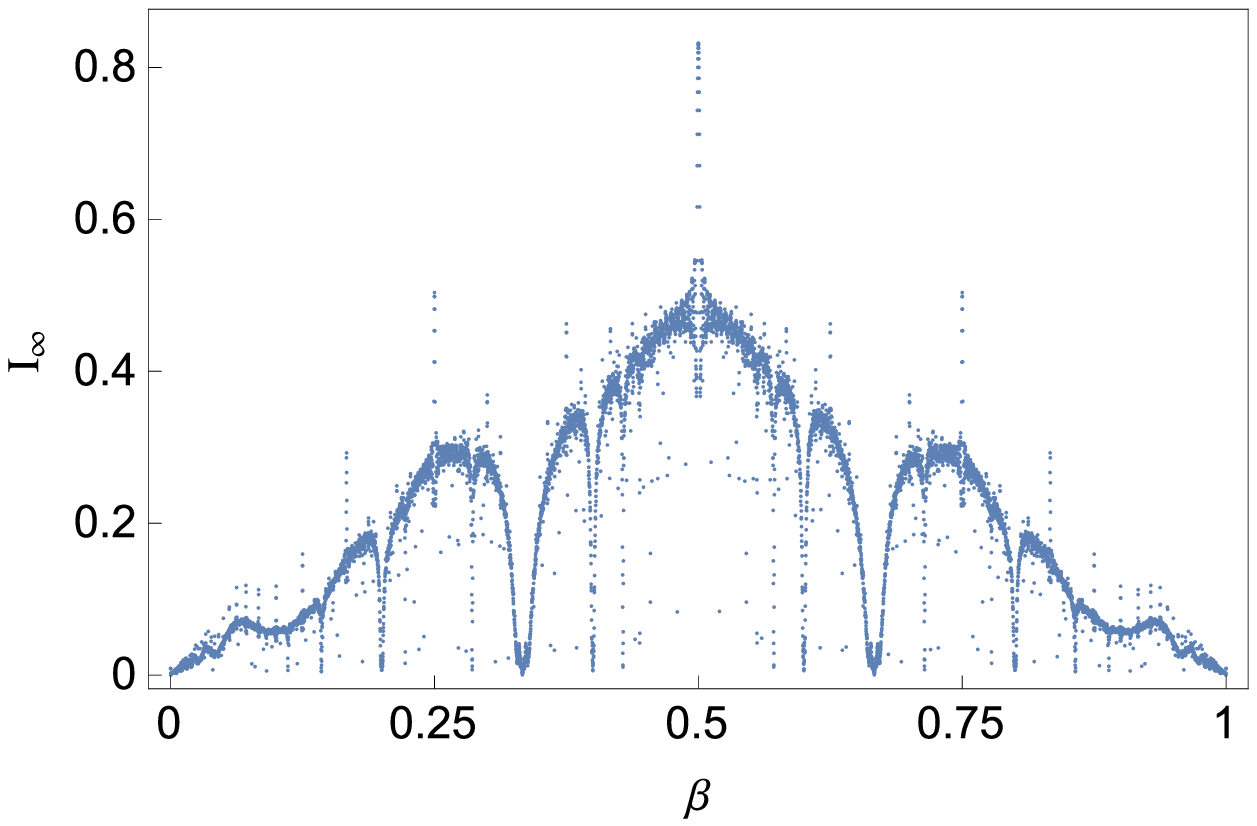}
\caption{(Color online) Long time density imbalance $I_\infty$ as a function of the period $\beta^{-1}$ of the superlattice (in units of the lattice constant for the primary lattice) for a noninteracting one dimensional system with $N_s=200$ sites and superlattice strength $\Delta/J=3$.} \label{fractalfig} 
\end{figure}

Fig.~\ref{fractalfig} shows $I_\infty$ as a function of $\beta$ where $N_s=200$, $\Delta/J=3$ and $\phi=0$. The behavior of the imbalance depends strongly on whether $\beta$ is irrational or rational, and thus displays a fractal structure. When $N_s \beta= N_s p/q$ is an integer, $I_\infty$ has peaks for even $q$ and troughs for odd $q$. Increasing $N_s$ leads to finer structure.

\subsection{Derivation of Cluster Expansion}
Here we will derive Eq.~(6) given in the main text. From Eq.~(4) we have 
\begin{equation} 
\begin{split}
\langle I(t)\rangle = & \frac{1}{Z} \Big[ \epsilon(1-\epsilon)^{N_s-1} \sum_{\{1\}} {}^{'} C_{\{1\}}(t)  + \frac{\epsilon^2}{2}(1-\epsilon)^{N_s-2} \sum_{\{2\}}  {}^{'}  C_{\{2\}}(t) \\
& +\frac{\epsilon^3}{3}(1-\epsilon)^{N_s-3} \sum_{\{3\}}  {}^{'}  C_{\{3\}}(t)  + ... + \frac{\epsilon^{N_s}}{N_s}   \sum_{\{N_s\}}  {}^{'}  C_{\{ N_s \} } (t) \Big].
\end{split}
\end{equation}
where $\{n\}= \{i_1 \sigma_1, i_2 \sigma_2, ..., i_n \sigma_n \}$ labels an $n$-particle initial state with particles at sites $i$ with spin $\sigma$, $\sum{}'_{\{n\}}$ denotes a sum over the $i_j$'s and $\sigma_j$'s such that the $i_j$'s are restricted to odd sites.

Substituting Eq.~(5) in Eq.~(4) we have

\begin{equation}
\begin{split}
 Z \langle I(t)\rangle= &  \epsilon(1-\epsilon)^{N_s-1} \sum_{\{1\}} {}^{'} \widetilde{C}_{\{1\}}(t)  + \frac{\epsilon^2}{2}(1-\epsilon)^{N_s-2} \sum_{\{2\}} {}^{'} \Big[ \widetilde{C}_{\{2\}}(t) +\sum_{  \langle\{1 \} \in \{2 \} \rangle } \widetilde{C}_{\{1\}}(t) \Big]\\ 
& +  \frac{\epsilon^3}{3}(1-\epsilon)^{N_s-3} \sum_{\{3\}} {}^{'}\Big[ \widetilde{C}_{\{3\}}(t) +\sum_{  \langle\{2 \} \in \{3 \} \rangle } \widetilde{C}_{\{2\}}(t) +\sum_{  \langle\{1 \} \in \{3 \} \rangle } \widetilde{C}_{\{1\}}(t)  \Big] + ...
\end{split}
\end{equation}
where $\sum_{  \langle\{k \} \in \{n \} \rangle }$ denotes a sum over all $\binom{n}{k}$ combinations of $k$ site and spin labels in $\{n\}$. For example, neglecting spin indices: $\sum_{\{2\}} {}^{'} \sum_{  \langle\{1 \} \in \{2 \}  \rangle } f( \{1 \})= \sum_{\substack{i_1 \rm{odd} \\ i_2 \rm{odd}}} ( f(i_1)+ f(i_2) )$.

We note the following identity:
\begin{equation}
\sum_{\{ n \} } {}^{'} \sum_{  \langle\{k \} \in \{n \} \rangle } \widetilde{C}_{\{k\}}(t) = \binom{N_s -k}{n-k} \sum_{ \{k\} } {}^{'} \widetilde{C}_{\{k\}} (t)
\end{equation}
 where the combinatorial  factor is the number of ways of choosing the $n-k$ elements of $\{n \}$ which are not in $\{ k \}$ out of the $N_s -k $ available starting positions/spins.

Substituting this identity into Eq.~(S3) yields

\begin{equation}
\begin{split}
 Z \langle I(t)\rangle = &  \epsilon(1-\epsilon)^{N_s-1} \sum_{\{1\}} {}^{'} \widetilde{C}_{\{1\}}(t) + \frac{\epsilon^2}{2}(1-\epsilon)^{N_s-2} \Big[ \sum_{\{2\}} {}^{'}  \widetilde{C}_{\{2\}}(t) + \binom{N_s-1}{1} \sum_{ \{1 \} } {}^{'}  \widetilde{C}_{\{1\}}(t) \Big] \\ 
& +  \frac{\epsilon^3}{3}(1-\epsilon)^{N_s-3}  \Big[ \sum_{\{3\}} {}^{'}   \widetilde{C}_{\{3\}}(t) +\binom{N_s-2}{1} \sum_{ \{2 \}  }  {}^{'} \widetilde{C}_{\{2\}}(t) + \binom{N_s-1}{2} \sum_{  \{1 \} } {}^{'} \widetilde{C}_{\{1\}}(t)  \Big]  + ...
\end{split}
\end{equation}

Collecting like terms, we have

\begin{equation}
\begin{split}
Z \langle I(t)\rangle& =  \sum^{N_s}_{n=1} \frac{1}{n} \epsilon^n (1-\epsilon)^{N_s-n} \binom{N_s-1}{n-1} \sum_{\{1\}} {}^{'} \widetilde{C}_{\{1\}}(t) \\ 
& +  \sum^{N_s}_{n=2} \frac{1}{n} \epsilon^n (1-\epsilon)^{N_s-n} \binom{N_s-2}{n-2} \sum_{\{2\}} {}^{'} \widetilde{C}_{\{2\}}(t) \\ 
& +  \sum^{N_s}_{n=3} \frac{1}{n} \epsilon^n (1-\epsilon)^{N_s-n} \binom{N_s-3}{n-3} \sum_{\{3\}} {}^{'} \widetilde{C}_{\{3\}}(t)\\
& + ... 
\end{split}
\end{equation}

which can be expressed as

\begin{equation} \label{eqisupp}
\begin{split}
\langle I(t)\rangle & =  A_1(\epsilon) \sum_{\{1\}} {}^{'} \widetilde{C}_{\{1\}}(t) + A_2(\epsilon) \sum_{\{2\}} {}^{'} \widetilde{C}_{\{2\}}(t) +  A_3(\epsilon) \sum_{\{3\}}{}^{'}  \widetilde{C}_{\{3\}}(t)+...+  \sum_{\{ N_s\}}{}^{'}  A_{N_s}(\epsilon) \widetilde{C}_{ \{N_s \}}(t)
\end{split}
\end{equation}
where $A_k(\epsilon) = \frac{1}{Z} \sum^{N_s}_{n=k} \frac{1}{n} \epsilon^n (1-\epsilon)^{N_s-n} \binom{N_s-k}{n-k} $. 
Taking the $N_s \to \infty$ limit gives Eq.~(6) to $O(\epsilon^2)$. Including finite size corrections, we have  $A_1(\epsilon)= \frac{1}{N_s}$ and $A_2(\epsilon)= \frac{\epsilon N_s -1 +(1-\epsilon)^{N_s}}{N_s(N_s-1)(1-(1-\epsilon)^{N_s})}$.

\subsection{Doublon Weighting}
Here we develop a cluster expansion for an ensemble averaged imbalance $\langle I(t) \rangle' $ which weights initial states with separate probabilities for doublons and singlons. We define $\langle I(t) \rangle' $ by 
\begin{equation} \label{doubeqmain}
\langle I(t) \rangle' = \frac{1}{\langle N \rangle Z} \sum_{\substack{n_\uparrow, n_\downarrow, n_d \\ n_\uparrow+n_\downarrow+n_d \leq N_s/2}} \sum{}' _{\{n_\uparrow \} \{n_\downarrow \} \{n_d \}} \rho^{n_\uparrow + n_\downarrow} \tau^{n_d} \langle \{n_\uparrow \} \{n_\downarrow \} \{n_d \} | \hat{n}_I (t) | \{n_\uparrow \} \{n_\downarrow \} \{n_d \} \rangle
\end{equation}
where $\{n_\uparrow \}= \{ i_1 , i_2 , ..., i_{n_\uparrow} \}$, $\{n_\downarrow \} =\{ j_1 , j_2 , ..., j_{n_\downarrow} \} $, and $\{n_d \}= \{ k_1 , k_2 , ..., k_{n_d} \}$ label the sites of up spin singlons, down spin singlons, and doublons, respectively. The symbol $\sum{}' _{\{n_\uparrow \} \{n_\downarrow \} \{n_d \}}$ denotes a sum over all possible locations of $n_\uparrow$ up-spin singlons, $n_\downarrow$ down-spin singlons, and $n_d$ doublons, restricted to odd sites. $\rho$ and $\tau$ are weights for the singlons and doublons. $Z$ is a normalization factor given by 
\begin{equation}
Z= \sum_{\substack{n_\uparrow, n_\downarrow, n_d \\ n_\uparrow+n_\downarrow+n_d \leq N_s/2}} \rho^{n_\uparrow+ n_\downarrow} \tau^{n_d} \binom{N_s/2}{n_\uparrow n_\downarrow n_d} = (1+2\rho+ \tau)^{N_s/2}
\end{equation}
where $\binom{N_s/2}{n_\uparrow n_\downarrow n_d} = \frac{N_s/2 !}{n_\uparrow ! n_\downarrow ! n_d ! (N_s/2- n_\uparrow n_\downarrow n_d)!}$ is the number of ways of assigning $n_\uparrow+ n_\downarrow$ singlons and $n_d$ doublons to $N_s/2$ odd sites. $\langle N \rangle$ is the mean number of particles and is given by 
\begin{equation} \label{avgdens}
\langle N \rangle = \frac{1}{Z}  \sum_{\substack{n_\uparrow, n_\downarrow, n_d \\ n_\uparrow+n_\downarrow+n_d \leq N_s/2}}  \rho^{n_\uparrow+ n_\downarrow} \tau^{n_d} \binom{N_s/2}{n_\uparrow n_\downarrow n_d} (n_\uparrow + n_\downarrow+ 2 n_d) = \frac{N_s (\rho+ \tau)}{1+2\rho+\tau}
\end{equation}

We define $C_{  \{n_\uparrow \} \{n_\downarrow \} \{n_d \}} (t)=  \langle \{n_\uparrow \} \{n_\downarrow \} \{n_d \} | \hat{n}_I (t) | \{n_\uparrow \} \{n_\downarrow \} \{n_d \} \rangle$. We decompose the expectation value $C(t)$ into single particle contributions, two particle contributions, etc. in a manner similar to Eq.~(5) in the main text:
\begin{equation} \label{doubdecomp}
\begin{split}
& C_{  \{n_\uparrow \} \{n_\downarrow \} \{n_d \}} (t)=  \sum_{ \langle \{1\} \in \{ n_\uparrow \} \rangle} C_{ \{1 \} \{0 \} \{0 \}} (t) + \sum_{ \langle \{1\} \in \{ n_\downarrow \} \rangle} C_{ \{0 \} \{1 \} \{0 \}} (t) \\ &+ \sum_{ \langle \{1 \} \in \{ n_d \} \rangle} C_{ \{0 \} \{0 \} \{1 \}} (t) + \sum_{ \langle \{1\} \in \{ n_\uparrow \}, \{1\} \in \{ n_\downarrow \} \rangle} \widetilde{C}_{ \{1 \} \{1 \} \{0 \}} (t) + ...
\end{split}
\end{equation}
$ \sum_{ \langle \{1\} \in \{ k\} \rangle} $ denotes a sum over all labels in $\{ k\}$ and $\widetilde{C}_{ \{1 \} \{1 \} \{0 \}} (t) = C_{\{1 \} \{1 \} \{0 \} } (t) - C_{\{1 \} \{0 \} \{0 \} } (t) -C_{\{0 \} \{1 \} \{0 \} } (t)$. There are higher particle number terms in this decomposition, but for the low density limit we consider here, it is sufficient (and notationally simpler) to keep terms up to two-body. We note that two-body terms like $\widetilde{C}_{ \{2\} \{0\} \{0 \}} (t) = C_{\{2 \} \{0 \} \{0 \} } (t) - C_{\{1 \} \{0 \} \{0 \} } (t) -C_{\{1 \} \{0 \} \{0 \} } (t)$ vanish, since two atoms with the same spin do not interact.

Substituting Eq.~(\ref{doubdecomp}) into Eq.~(\ref{doubeqmain}) and performing simple summations yields
\begin{equation} \label{doubfinal}
\begin{split}
&\langle I(t) \rangle' = \frac{\rho}{\rho+\tau} \frac{1}{N_s} \left( \sum_{ \{1\} } {}' C_{ \{1 \} \{0 \} \{0 \}} (t)  + \sum_{\{1\} } {}' C_{ \{0 \} \{1 \} \{0 \}} (t) \right)  \\
&+ \frac{\tau}{\rho+\tau} \frac{1}{N_s} \sum_{ \{1 \} } {}' C_{ \{0 \} \{0 \} \{1 \}} (t) + \frac{\rho+\tau}{(1+2\rho +\tau)} \frac{1}{N_s} \sum{}' _{  \{1\} , \{1\} } \widetilde{C}_{ \{1 \} \{1 \} \{0 \}} (t)
\end{split}
\end{equation}

We vary $\rho$ and $\tau$ in Eq.~(\ref{doubfinal}) to produce Fig.~3 in the main text.

\subsection{Local Integrals of Motion}

Fig.~S2 shows the coefficients $U^{(2)}_{\substack{mn\\ \uparrow \downarrow}}$ which appear in Eq.~(10) of the main text.

\begin{figure}
\includegraphics[width=0.5\textwidth]{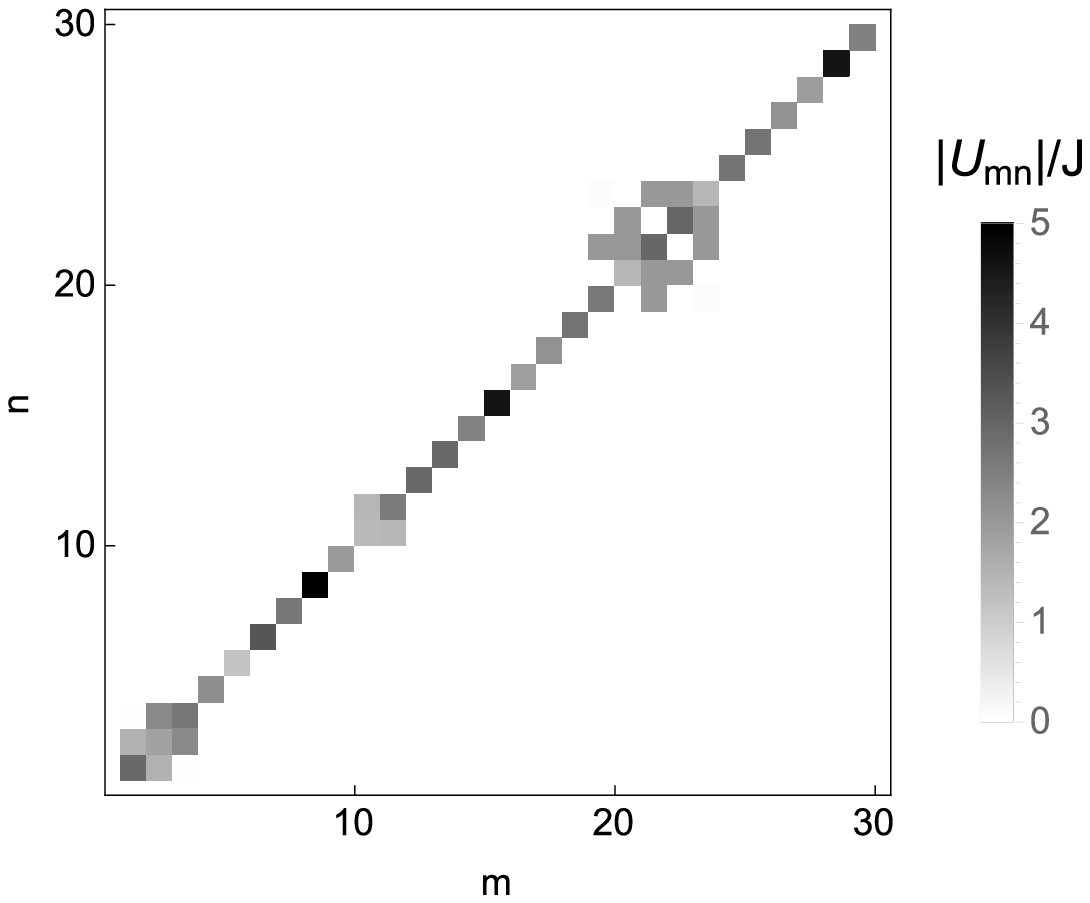}
\caption{Two particle interaction term $U^{(2)}_{{mn\\ \uparrow \downarrow}}$ appearing in Eq.~(10) of the main text. Here $\Delta/J=3$, $U/J= 3$, and $\beta=0.721$. Darker colors correspond to larger values of $|U^{(2)}_{mn}|$. For large $|n-m|$, $U^{(2)}_{nm}$ is exponentially small. For $n=m$, $U^{(2)}_{mn} \sim U$.} 
\end{figure}

\end{document}